\documentclass[twocolumn,flushbottom,noshowpacs,preprintnumbers,superscriptaddress,amsmath,amssymb]{revtex4}
\usepackage{graphicx}% Include figure files
\usepackage{dcolumn}% Align table columns on decimal point
\usepackage{bm}% bold math
\usepackage{amssymb}
\usepackage{amssymb}
\def\>{\rangle}

\begin{document}

\title{Giant nonlinearity via breaking parity-time symmetry: a route to low-threshold phonon diodes}% Force line breaks with \\

\author{Jing Zhang}\email{jing-zhang@mail.tsinghua.edu.cn}
\affiliation{Department of Automation, Tsinghua University,
Beijing 100084, P. R. China} \affiliation{Center for Quantum
Information Science and Technology, TNList, Beijing 100084, P. R.
China} \affiliation{Department of Electrical and Systems
Engineering, Washington University, St. Louis, MO 63130, USA}
\author{Bo Peng}\affiliation{Department of Electrical and Systems
Engineering, Washington University, St. Louis, MO 63130, USA}
\author{\c{S}ahin Kaya \"{O}zdemir}\email{ozdemir@ese.wustl.edu}
\affiliation{Department of Electrical and Systems Engineering,
Washington University, St. Louis, MO 63130, USA}
\author{Yu-xi Liu}\email{yuxiliu@mail.tsinghua.edu.cn}
\affiliation{Institute of Microelectronics, Tsinghua University,
Beijing 100084, P. R. China} \affiliation{Center for Quantum
Information Science and Technology, TNList, Beijing 100084, P. R.
China}
\author{Hui Jing}\affiliation{Department of Physics, Henan Normal University, Xinxiang
453007, P. R. China}\affiliation{CEMS, RIKEN, Saitama 351-0198,
Japan}
\author{Xin-you L\"{u}}\affiliation{School of physics, Huazhong University of Science and
Technology, Wuhan 430074, China}\affiliation{CEMS, RIKEN, Saitama
351-0198, Japan}
\author{Yu-long Liu}\affiliation{Institute of Microelectronics, Tsinghua University,
Beijing 100084, P. R. China}
\author{Lan Yang}\email{yang@ese.wustl.edu}\affiliation{Department of Electrical and Systems Engineering, Washington University, St. Louis, MO 63130, USA}
\author{Franco Nori}\email{fnori@riken.jp}\affiliation{CEMS, RIKEN, Saitama 351-0198,
Japan}\affiliation{Physics Department, The University of Michigan,
Ann Arbor, MI 48109-1040, USA}

\date{\today}% It is always \today, today,
             %  but any date may be explicitly specified

\begin{abstract}
Nonreciprocal devices that permit wave transmission in only one
direction are indispensible in many fields of science including,
e.g., electronics, optics, acoustics, and thermodynamics.
Manipulating phonons using such nonreciprocal devices may have a
range of applications such as phonon diodes, transistors,
switches, etc. One way of achieving nonreciprocal phononic devices
is to use materials with strong nonlinear response to phonons.
However, it is not easy to obtain the required strong mechanical
nonlinearity, especially for few-phonon situations. Here, we
present a general mechanism to amplify nonlinearity using
$\mathcal{PT}$-symmetric structures, and show that an on-chip
micro-scale phonon diode can be fabricated using a
$\mathcal{PT}$-symmetric mechanical system, in which a lossy
mechanical-resonator with very weak mechanical nonlinearity is
coupled to a mechanical resonator with mechanical gain but no
mechanical nonlinearity. When this coupled system transits from
the $\mathcal{PT}$-symmetric regime to the
broken-$\mathcal{PT}$-symmetric regime, the mechanical
nonlinearity is transferred from the lossy resonator to the one
with gain, and the effective nonlinearity of the system is
significantly enhanced. This enhanced mechanical nonlinearity is
almost lossless because of the gain-loss balance induced by the
$\mathcal{PT}$-symmetric structure. Such an enhanced lossless
mechanical nonlinearity is then used to control the direction of
phonon propagation, and can greatly decrease (by over three orders
of magnitude) the threshold of the input-field intensity necessary
to observe the unidirectional phonon transport. We propose an
experimentally realizable lossless low-threshold phonon diode of
this type. Our study opens up new perspectives for constructing
on-chip few-phonon devices and hybrid phonon-photon components.
\end{abstract}


\maketitle

\section{Introduction}\label{s1}
Owing to recent progress in nanotechnology and materials science,
nano- and
micro-mechanics~\cite{TJKippenbergScience:2008,MPootPR:2012,MBlencowePR:2004,YaSGreenbergPU:2012,MAspelmeyerRMP:2014,NLiRMP:2012,LWangPRL2008}
have emerged as subjects of great interest due to their potential
use in demonstrating macroscopic quantum phenomena, and possible
applications in precision measurements, detecting gravitational
waves, building filters, signal amplification, as well as switches
and logic gates. In particular, on-chip single- or few-phonon
devices are ideal candidates for hybrid quantum information
processing, due to the ability of phonons to interact and rapidly
switch between optical fields and microwave
fields~\cite{RWAndrewsNatPhys:2014,JBochmannNatPhys:2013}.
Fabrication of high-frequency mechanical
resonators~\cite{ADOConnellNature:2010}, demonstration of coherent
phonon coupling between nanomechanical
resonators~\cite{QLinNatPhoton:2010}, ground-state
cooling~\cite{JChanNature:2011,JDTeufelNature:2011}, optomechanics
(in microtoroids~\cite{TJKippenbergPRL:2005,TCarmonPRL:2005},
microspheres~\cite{YSPark,CDongScience:2012,MTomesPRL:2009},
microdisks~\cite{JHoferPRA:2010,LDingAPL:2011,MZhangAPL:2014},
microring~\cite{LFanNatCommu:2015}, photonic
crystals~\cite{QLinNatPhoton:2010}, doubly- or singly-clamped
cantilevers~\cite{CHMetzgeNature:2004,OArcizetNature:2006}, and
membranes~\cite{JDThompson}) have opened new
directions~\cite{MAspelmeyerRMP:2014} and provided new tools to
control and manipulate phonons in on-chip devices. One possible
obstacle to further develop this field is the ability to control
the flow of phonons, allowing transport in one direction but not
the opposite direction~\cite{RFleuryScience:2014}, i.e.,
nonreciprocal phonon transport. There have been several attempts
to fabricate nonreciprocal devices for
phonons~\cite{AAMaznevWM:2007,RKrishnanSSC:2007,XFLiPRL:2011,BLiangNatMat:2010,NBoechlerNatMat:2011,BIPopaNatCommu:2014},
but these are almost exclusively based on asymmetric linear
structures which indeed cannot break Lorentz reciprocity: a static
linear structure cannot break reciprocity~\cite{AAMaznevWM:2007}.
These proposed linear structures do obey the
reflection-transmission reciprocity and thus cannot be considered
as ``phonon diodes". Diode-like behavior was observed in these
linear acoustic structures because the input-output channels were
not properly switched~\cite{AAMaznevWM:2007}.

Nonreciprocal phonon transmission inevitably requires
magneto-acoustic materials, strong nonlinearity, or a
time-dependent modulation of the parameters of a structure.
Although already demonstrated in optics~\cite{KPetermann:1988},
the time-dependent modulation of acoustic parameters of a phononic
structure has not been probed yet. Magneto-acoustic materials
require high magnetic fields to operate and have been
studied~\cite{SRSklanNJP:2014}; however, a magnetic-free
nonreciprocal device is critical for building on-chip and
small-scale phononic processors and circuits. Nonlinearity-based
nonreciprocity seems to be the most viable approach for creating
micro- or nano-scale nonreciprocal devices for controlling and
manipulating phonons.

Recently, there have been several reports on nonlinear mechanical
structures and
materials~\cite{RAlmogPRL:2007,MHMathenyPRL:2014,KLEkinciPSI:2005,IMabhoobNatCommun:2011}.
However, the weak nonlinearity of those acoustic/phononic
materials hinders progress in this direction due to the high input
powers required to observe the nonlinear
effects~\cite{ANCleland:2003,XYLvPRA:2014}. In order to circumvent
this problem, coupling a weakly nonlinear structure to an
auxiliary system, such as a quantum bit~\cite{KJacobsPRL:2009},
has been proposed to engineer effective giant mechanical
nonlinearities.

In order to achieve the required nonlinearity for nonreciprocal
phonon transport and to study nonlinear phononics, here we
introduce a new method based on parity-time ($\mathcal{PT}$)
symmetry~\cite{CMBenderRPP:2007}, which has attracted much
attention recently due to their interesting and generally
counter-intuitive
physics~\cite{AASukhorukovPRA:2010,HRamezaniPRA:2010,ZLinPRL:2011,XZhuOL:2013,CHangPRL:2013,GSAgarwalPRA:2012,HBenistyOE:2011,NLazaridesPRL:2013,YLumerPRL:2013,AGuoPRL:2009,CERuterNatPhys:2010,LFengScience:2011,ARegensburgerNature:2012,LFengNatMat:2012,JSchindlerPRAR:2011,CZhengPTRSA:2013,SBittnerPRL:2012,CMBenderAJP:2013,NBenderPRL:2013,NMChtchelkatchev,BPengNatPhys2014,LChangNatPhoton:2014,XFZhuPRX:2014,RFleuryNatCommu:2015,HJingPRL:2014,XWXu}.
Parity-time symmetry and its breaking (broken
$\mathcal{PT}$-phase) have been demonstrated in various physical
systems~\cite{AGuoPRL:2009,CERuterNatPhys:2010,LFengScience:2011,ARegensburgerNature:2012,LFengNatMat:2012,JSchindlerPRAR:2011,CZhengPTRSA:2013,SBittnerPRL:2012,CMBenderAJP:2013,NBenderPRL:2013,NMChtchelkatchev,BPengNatPhys2014,LChangNatPhoton:2014},
such as optical
waveguides~\cite{AGuoPRL:2009,CERuterNatPhys:2010,LFengScience:2011,ARegensburgerNature:2012},
microcavities~\cite{BPengNatPhys2014}, and electrical
circuits~\cite{JSchindlerPRAR:2011}. However, mechanical
$\mathcal{PT}$-symmetric systems have only been considered quite
recently~\cite{CMBenderAJP:2013,XFZhuPRX:2014,RFleuryNatCommu:2015,HJingPRL:2014,XWXu}.

In our proposed mechanical $\mathcal{PT}$ symmetric system, a
lossy mechanical resonator (passive resonator) which has a weak
mechanical nonlinearity is coupled to a mechanical resonator with
mechanical gain (active resonator) that balances the loss of the
passive resonator. The active resonator here works as a dynamical
amplifier. In the vicinity of the $\mathcal{PT}$-phase transition,
the weak nonlinearity is first distributed between the
mechanically-coupled resonators and then significantly enhanced
due to the localization of the mechanical supermodes in the active
resonator. In this way, the effective nonlinear Kerr coefficient
is increased by over three orders of magnitude. This strong
nonlinearity, localized in the active resonator, blocks the phonon
transport from the active resonator to the lossy resonator but
permits the transport in the opposite direction.

For the experimental realization of the proposed
nonlinearity-based phonon diode, we provide a system in which a
mechanical beam with weak mechanical nonlinearity is coupled to
another mechanical beam with gain. We show that this micro-scale
system can be switched from a bidirectional transport regime to a
unidirectional transport regime, and vice versa, by properly
adjusting the detuning between the mechanical frequency of the
resonators and the frequency of the driving phononic field, or by
varying the amplitude of the input phononic field.

\section{Parity-time ($\mathcal{PT}$-) symmetric mechanical system}\label{s2}
\begin{figure} \centerline{\includegraphics[width =
8.6 cm]{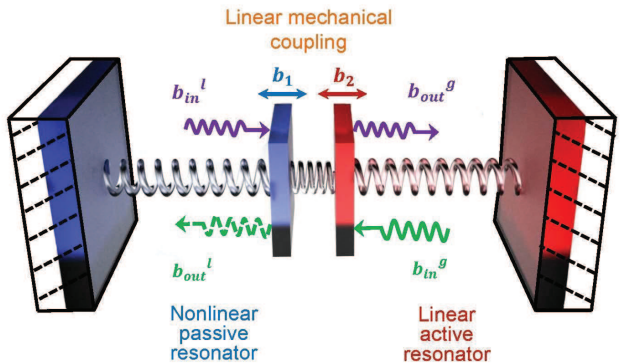}}\caption{(Color online) Schematic diagram of the
proposed $\mathcal{PT}$-symmetric mechanical system. The
$\mathcal{PT}$-symmetric mechanical system has a linear mechanical
coupling between a passive mechanical resonator (having mechanical
loss and very weak mechanical nonlinearity) and an active
mechanical resonator (having mechanical gain but no nonlinearity).
Here $b_{in}^l$ and $b_{in}^g$ are the input fields to the passive
and active resonators, respectively, and $b_{out}^l$ and
$b_{out}^g$ are the output fields, respectively, leaving the
passive and active resonators. $b_1$ and $b_2$ denote the movable
resonators.}\label{Fig of Mechanical PT}
\end{figure}

The system we consider here consists of two mechanical resonators,
one of which has mechanical loss (passive resonator) and weak
nonlinearity, and the other has mechanical gain (active resonator)
but no nonlinearity (see Fig.~\ref{Fig of Mechanical PT}). The
mechanical coupling between the resonators is linear and it gives
rise to the mechanical supermodes $b_{\pm}$ with complex
eigenfrequencies
\begin{equation}\label{Complex frequencies for the PT system}
\omega_{\pm}=\Omega_{\pm}-i\Gamma_{\pm},
\end{equation}
given by
\begin{equation}\label{Complex frequencies for the PT system with additional terms}
\omega_{\pm}=\Omega_0-i\chi\pm\beta.
\end{equation}
Here $\Omega_0$ is the mechanical frequency of the solitary
mechanical resonators (i.e., both resonators are degenerate),
\begin{eqnarray}\label{System parameters}
&\chi=\left(\Gamma_l-\Gamma_g\right)/2,&\\
&\beta=\sqrt{g_{mm}^2-\Gamma^2},&\\
&\Gamma=\left(\Gamma_l+\Gamma_g\right)/2,&
\end{eqnarray}
where $\Gamma_l$ and $\Gamma_g$ denote, respectively, the damping
rate of the lossy mechanical resonator and the gain rate of the
active mechanical resonator, and $g_{mm}$ is the coupling strength
between the mechanical modes. When $\Gamma\leq g_{mm}$, the system
is in the $\mathcal{PT}$-symmetric regime, and the supermodes are
non-degenerate with
\begin{equation}\label{Split mode}
\Omega_{\pm}=\Omega_0\pm\beta
\end{equation}
and have the same damping rate $\chi$ (see Figs.~\ref{Fig of
amplifying mechanical nonlinearity}a and ~\ref{Fig of amplifying
mechanical nonlinearity}b). However, when $\Gamma>g_{mm}$, the
system is in the broken-$\mathcal{PT}$-symmetric regime, the
supermodes are frequency-degenerate with $\Omega_{\pm}=\Omega_0$
(see Figs.~\ref{Fig of amplifying mechanical nonlinearity}a and
~\ref{Fig of amplifying mechanical nonlinearity}b) and have
different damping rates
\begin{equation}\label{Damping rate}
\Gamma_{\pm}=\chi\mp i\beta.
\end{equation}
At $\Gamma=g_{mm}$, the
two supermodes are degenerate with the same damping rate,
indicating a transition between the PT-symmetric regime and the
broken-PT-symmetry regime. This point is generally referred to as
the $\mathcal{PT}$-transition point. It is seen that the two
supermodes will be lossless in the $\mathcal{PT}$-symmetric regime
if the gain and loss are well-balanced, such that
$\Gamma_l=\Gamma_g$.

\section{Enhancing mechanical nonlinearity by breaking $\mathcal{PT}$-symmetry}\label{s3}

Let us assume that the passive resonator is made from a nonlinear
acoustic material~\cite{RAlmogPRL:2007} with a small nonlinear
Kerr coefficient $\mu$. This nonlinearity mediates a cross-Kerr
interaction between the two mechanical supermodes, which leads to
the effective nonlinear coefficients $\mu_b'$ and $\mu_s'$, in the
broken- and unbroken-$\mathcal{PT}$ regimes~\cite{SI}:
\begin{equation}\label{Enhanced nonlinear coefficient}
\mu_b'=\mu\frac{\Gamma^2
g_{mm}^2}{\left(\Gamma^2-g_{mm}^2\right)^2},\quad\mu_s'=\mu\frac{g_{mm}^4}{\left(\Gamma^2-g_{mm}^2\right)^2}.
\end{equation}
Clearly, the effective nonlinear coefficients are significantly
enhanced in the vicinity of the phase transition point
$\Gamma=g_{mm}$. Moreover, if the gain and loss are well-balanced,
i.e., $\Gamma_l=\Gamma_g$, the supermodes become almost lossless.
This observation is one of the key contributions of this paper.
Namely, operating the system of two coupled mechanical resonators
in the vicinity of the phase transition point will significantly
enhance the existing very weak nonlinearity with an extremely
small loss rate.
\begin{figure} \centerline{\includegraphics[width =
6.8 cm]{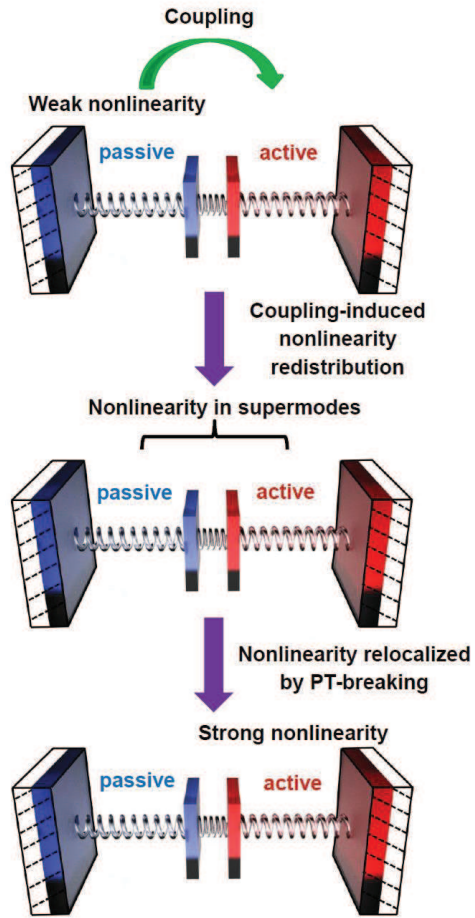}}\caption{(Color online) Enhancement of
mechanical nonlinearity in a PT-symmetric mechanical system. The
coupling between two mechanical resonators creates two mechanical
supermodes symmetrically distributed between the resonators, and
hence both supermodes experience the weak nonlinearity of the
passive resonator. In the vicinity of the PT-phase transition,
which takes place when the coupling strength between the
resonators equals to the total loss in the system, the mechanical
nonlinearity is significantly enhanced due to localization of the
mechanical supermodes in the active mechanical
resonator.}\label{Fig of nonlinearity transfer}
\end{figure}

Using the parameter values of $\mu/\Omega_0=10^{-5}$,
$\Gamma_l/\Omega_0=0.55\times10^{-3}$, and
$\Gamma_g/\Omega_0=0.45\times10^{-3}$, we show in Fig.~\ref{Fig of
amplifying mechanical nonlinearity}  the evolution of the
eigenfrequencies of the system and of the nonlinear coefficient as
a function of $g_{mm}/\Gamma$. The transition from the broken- to
the unbroken-$\mathcal{PT}$ symmetric regime and vice versa, as
the mechanical coupling strength is varied, is seen in Fig.
~\ref{Fig of amplifying mechanical nonlinearity}a and \ref{Fig of
amplifying mechanical nonlinearity}b and it is reflected in the
bifurcations of the supermode frequencies and damping rates.
Moreover, the enhancement of the nonlinearity in the vicinity of
the $\mathcal{PT}$-phase transition point is seen in Fig.~\ref{Fig
of amplifying mechanical nonlinearity}c. We find that the
nonlinear coefficient is enhanced by more than three orders of
magnitude in the vicinity of the transition point.

More interestingly, in the broken-$\mathcal{PT}$ regime, the
mechanical energy of the coupled system is localized in the active
resonator, which leads to a nonlinear mechanical mode with strong
self-Kerr nonlinearity localized in the active mechanical
resonator. This can be interpreted intuitively as follows. The
initial weak mechanical nonlinearity is transferred from the
passive resonator to the active resonator and it is enhanced by
field localization in the broken-$\mathcal{PT}$ regime. Owing to
the presence of the mechanical gain, the active resonator then
enjoys an almost lossless mechanical mode with a giant
nonlinearity (see Fig.~\ref{Fig of nonlinearity transfer}).

Finally, we would like to consider how the mechanical nonlinearity
will affect the $\mathcal{PT}$-symmetric structure of the system.
Generally speaking, a strong nonlinearity will shift the
transition point of a $\mathcal{PT}$-symmetric system or even
destroy the $\mathcal{PT}$ symmetry of such a
system~\cite{IVBarashenkovPRA:2014}. However, in our case, we
start from a system in which a gain resonator is coupled to a
lossy resonator with very weak Kerr nonlinearity, and thus we can
omit the shift of the $\mathcal{PT}$-transition point induced by
such a weak nonlinearity. Although we generate a strong
nonlinearity in the vicinity of the $\mathcal{PT}$-transition
point, this is an effective nonlinearity induced in the supermode
picture and thus will not affect the supermodes and the
$\mathcal{PT}$-transition point of the system.
\begin{figure*}[t]\centerline{\includegraphics[width=16
cm]{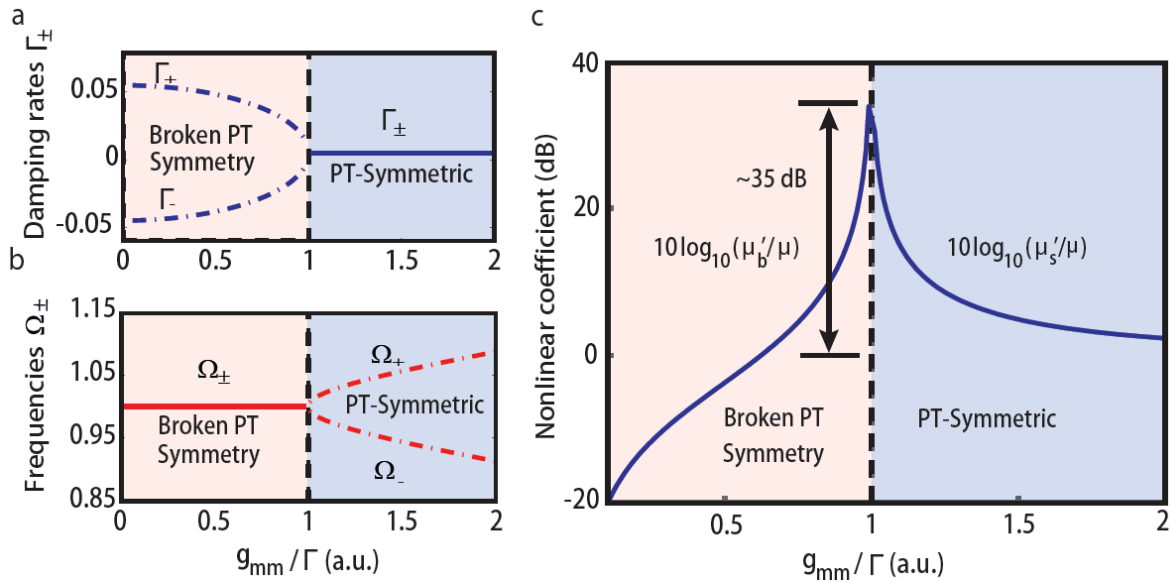}}\caption{(Color online) Amplification of mechanical
nonlinearity via $\mathcal{PT}$-symmetry breaking. (a) Effective
damping rates and (b) frequencies of the mechanical supermodes as
functions of the normalized mechanical coupling strength
$g_{mm}/\Gamma$. (c) The effective nonlinear coefficients $\mu_b'$
in the $\mathcal{PT}$-breaking regime and $\mu_s'$ in the
$\mathcal{PT}$-symmetric regime. The $\mathcal{PT}$-phase
transition takes place at $g_{mm}=\Gamma$. In the vicinity of this
transition point, the nonlinear coefficients $\mu_b'$ and $\mu_s'$
are enhanced by more than three orders of magnitude (more than
$35$ dB increase compared to the baseline).}\label{Fig of
amplifying mechanical nonlinearity}
\end{figure*}

\section{Unidirectional phonon transport}\label{s4}
Here we investigate the effect of the enhanced mechanical
nonlinearity on the phonon transport in the coupled system. We
find that the localized strong mechanical nonlinearity leads to
unidirectional phonon transport from the passive resonator to the
active resonator and blocks phonon transport in the opposite
direction (i.e., phonon transport from the active to the passive
resonator is prevented). The transport is almost lossless due to
the gain-loss balance of the system.  When this system is operated
in the vicinity of the $\mathcal{PT}$-phase transition point, the
unidirectional phonon transport is possible within a region given
by~\cite{SI}
\begin{equation}\label{Nonreciprocal window for detuning}
\delta\in\left[\frac{g_{mm}^2\Omega_0}{\Omega_0^2+\chi^2},\
\frac{g_{mm}^2}{\Omega_0-\sqrt{3}\chi}\right],
\end{equation}
where
\begin{equation}\label{Detuning between the frequencies of the resonator and the driving field}
\delta=\Omega_0-\Omega_d
\end{equation}
is the detuning between the input (driving) field frequency
$\Omega_d$ and the resonance frequency $\Omega_0$ of the
mechanical resonators. Additionally, in order to observe the
unidirectional phonon transport, the amplitude of the input field
should satisfy
\begin{equation}\label{Nonreciprocal window for input intensity}
\left|\varepsilon_d\right|^2\in\left[\frac{2\left(\delta^2+g_{mm}^2\right)^3}{9\mu_b'\delta^3},\
\frac{2\left(\delta^2+g_{mm}^2\right)^3}{9\mu_b'g_{mm}^2\delta}\right],
\end{equation}
implying that the intensity of the input field required for
unidirectional transport is inversely proportional to the strength
of the mechanical nonlinearity $\mu_b'$. Since the strength of the
mechanical nonlinearity can be enhanced by more than three orders
of magnitude by breaking the $\mathcal{PT}$ symmetry, the
threshold of the input-field intensity for observing
unidirectional phonon transport can be decreased by at least three
orders of magnitude, allowing a low-threshold phonon diode
operation.

\begin{figure} \centerline{\includegraphics[width =
7.0 cm]{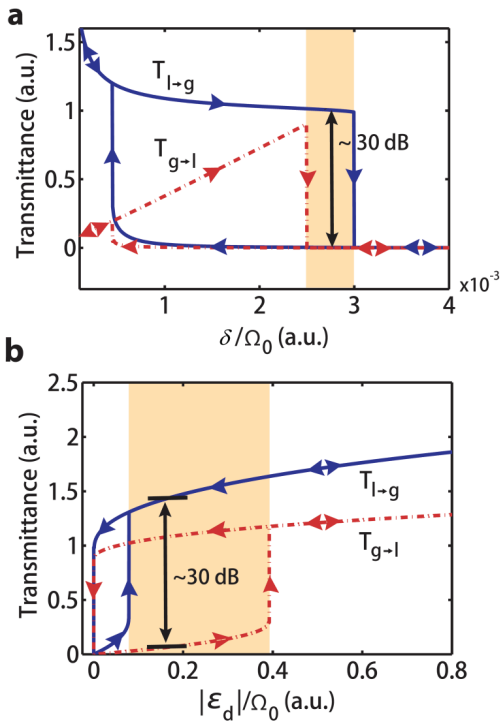}}\caption{(Color online) Unidirectional phonon
transport by $\mathcal{PT}$-symmetry breaking. (a) Unidirectional
phonon transport when the detuning $\delta$ is varied. The
transmittance from the active to passive mechanical resonator
$T_{g\rightarrow l}$ (red dash-dotted curve), and from the passive
to the active mechanical resonator $T_{l\rightarrow g}$ (blue
solid curve) versus the detuning $\delta=\Omega_0-\Omega_d$ shows
a strong bistability and hysteresis effect. The transmittance
functions evolve along different trajectories for increasing and
decreasing detuning due to the nonlinearity-induced bistability. A
unidirectional phonon-transport region (melon-colored shaded
region) appears only when the detuning $\delta$ is up-scanned from
smaller to larger detunings. Within this regime, the rectification
is $\sim\,30$ dB. (b) Unidirectional photon transport when the
amplitude of the input field is varied at fixed detuning
$\delta/\Omega_0=2.75\times10^{-3}$. Within the unidirectional
transport region (melon-colored shaded region), rectification is
$\sim\,30$ dB.}\label{Fig of unidirectional phonon transport}
\end{figure}

To show unidirectional phonon transport in the
broken-$\mathcal{PT}$ regime, let us first fix the amplitude of
the input field and vary the detuning $\delta$. We compare the
amplitude transmittance
\begin{equation}\label{Amplitude transmittance from loss to gain resonators}
t_{l\rightarrow g}=b_{\rm out}^g/b_{\rm in}^l
\end{equation}
and
\begin{equation}\label{Amplitude transmittance from gain to loss resonator}
t_{g\rightarrow l}=b_{\rm out}^l/b_{\rm in}^g.
\end{equation}
The former, $t_{l\rightarrow g}$, denotes the transmission from
the passive to the active resonator, that is, the system is driven
by a phononic input field $b_{\rm in}^l$ of frequency $\Omega_d$
at the passive resonator side and the output $b_{\rm out}^g$ is
measured at the active resonator side. However, the latter,
$t_{g\rightarrow l}$, denotes the amplitude transmittance from the
active resonator to the passive resonator when the system is
driven by the field $b_{\rm in}^g$ of frequency $\Omega_d$ at the
active resonator side and the output $b_{\rm out}^l$ is measured
at the passive side. The nonlinearity in the system manifests as a
bistability and hysteresis in the power transmittance,
\begin{equation}\label{Power transmittance from gain to loss resonators}
T_{g\rightarrow l}=\left|t_{g\rightarrow l}\right|^2
\end{equation}
and
\begin{equation}
T_{l\rightarrow g}=\left| t_{l\rightarrow g} \right|^2,
\end{equation}
obtained as the detuning $\delta$ is up-scanned from smaller to
larger detuning and down-scanned from larger to smaller detuning
(see Fig.~\ref{Fig of unidirectional phonon transport}a).

We find that during the down-scan, both of the transmittances
$T_{l\rightarrow g}$ and $T_{g\rightarrow l}$ stay at the lower
branch with values close to zero until $\delta/\Omega_0=0.5\times
10^{-3}$, after which they bifurcate from each other only slightly
and then jump to the stable points at the upper branch of their
respective trajectories (see Fig.~\ref{Fig of unidirectional
phonon transport}a). Further decreasing the detuning leads to an
increase in $T_{l\rightarrow g}$, but a decrease in
$T_{g\rightarrow l}$. This implies that there is no unidirectional
phonon transport with the parameter values used in the numerical
simulations. Instead, when the detuning is below a critical value,
the phonon transport is bidirectional; whereas when it is above
that critical value there is no phonon transport.

During the up-scan, however, after a short stay on the stable
state, i.e., a regime in which there is no bistability and
hysteresis in the transmittance, (during which $T_{l\rightarrow
g}$ decreases and $T_{g\rightarrow l}$ increases with growing
detuning), both of the transmittances follow the upper branches of
their trajectories, during which a linear increase in
$T_{g\rightarrow l}$ and a slow-rate decrease in $T_{l\rightarrow
g}$ are observed (see Fig.~\ref{Fig of unidirectional phonon
transport}a). This behavior continues until
$\delta/\Omega_0\sim2.5\times 10^{-3}$ for $T_{g\rightarrow l}$,
where it jumps to the lower branch of its trajectory, and becomes
zero as the detuning is increased (see Fig.~\ref{Fig of
unidirectional phonon transport}a). This implies that phonon
transport from the active mechanical resonator to the passive one
is prevented if the detuning is set to $\delta/\Omega_0>2.5\times
10^{-3}$. The transmittance $T_{l\rightarrow g}$ stays at its
upper branch with a value close to one until $\delta/\Omega_0\sim
3\times 10^{-3}$, where it jumps to its lower branch and becomes
zero. Thus, for $\delta/\Omega_0>3\times 10^{-3}$, phonon
transport from the passive to the active resonator is prevented.
Clearly, in the detuning region $2.5\times
10^{-3}<\delta/\Omega_0<3\times 10^{-3}$, the transmittance
$T_{l\rightarrow g}$ is close to one whereas $T_{g\rightarrow l}$
is close to zero in this detuning region phonon transport from the
active mechanical resonator to the passive one is forbidden,
whereas phonon transport from the passive mechanical resonator to
active one is allowed with almost no loss. Thus, we conclude that
phonon transmission is non-reciprocal in this detuning region, and
the rectification is $\sim\,30$ dB within the nonreciprocal
transport region (see Fig.~\ref{Fig of unidirectional phonon
transport}a). For detuning values smaller than the lower bound of
this region, phonon transport is bidirectional. For detuning
values larger than the upper bound of the region, phonon transport
is not possible.

Note that our phonon diode should work only when the disturbance
and perturbation of the system parameters are not too strong. In
fact, within the unidirectional phonon transport window shown in
Fig.~\ref{Fig of unidirectional phonon transport}a, the
transmittance $T_{l\rightarrow g}$ has two different branches of
metastable values. When we increase the detuning $\delta$ within
this unidirectional phonon transport window, $T_{l\rightarrow g}$
will stay in the upper stable branch if we do not severely disturb
the system and the phonon diode should operate properly. However,
if the disturbance is too strong, $T_{l\rightarrow g}$ will jump
from the upper branch to the lower branch and stay in this stable
lower branch, without rectification.

Alternatively, we can fix the detuning and vary the amplitude of
the input field to show the nonlinearity-induced bistability and
hysteresis. A nonreciprocal phonon transport region is seen when
the amplitude of the input field is up-scanned (see Fig.~\ref{Fig
of unidirectional phonon transport}b). The nonreciprocal transport
region disappears when the amplitude of the input field is
down-scanned. Within the nonreciprocal transport region, when the
input is varied at fixed detuning (see Fig.~\ref{Fig of
unidirectional phonon transport}b), the rectification is
$\sim\,30$ dB. Similarly, in this case, due to the metastability
of the transmittance $T_{l\rightarrow g}$, the disturbance-induced
perturbation of the system parameters may not be too strong
otherwise our design of phonon diode will be invalid.

\section{On-chip phonon diode}\label{s5}
The unidirectional phonon transport enabled by the
$\mathcal{PT}$-breaking-induced strong mechanical nonlinearity can
be used to fabricate lossless phonon diodes in on-chip systems.
This may have many applications, such as single-phonon transistors
and routers, on-chip quantum switches, and information-processing
components. One possible way to realize the proposed phonon diode
is to use coupled beams and cantilevers (see Fig.~\ref{Fig of
experimental setup}a). Phonon lasing, and hence an active
mechanical resonator, has been experimentally realized in an
electromechanical beam~\cite{IMahboobPRL:2013}.
Elastically-coupled nano beams and cantilevers, by which the
mechanical supermodes can be generated, have also been shown in
various
experiments~\cite{EGilSantosNanoLetters:2009,RBKarabalinPRB:2009,HOkamotoAPE:2009,TSBiswasNanoLetters:2014},
in which the two mechanical resonators can be independently
driven~\cite{RBKarabalinPRB:2009}. Thus our proposal is within the
reach of current experimental techniques of
nano-micro-electromechanical systems.

\begin{figure} \centerline{\includegraphics[width =
8.6 cm]{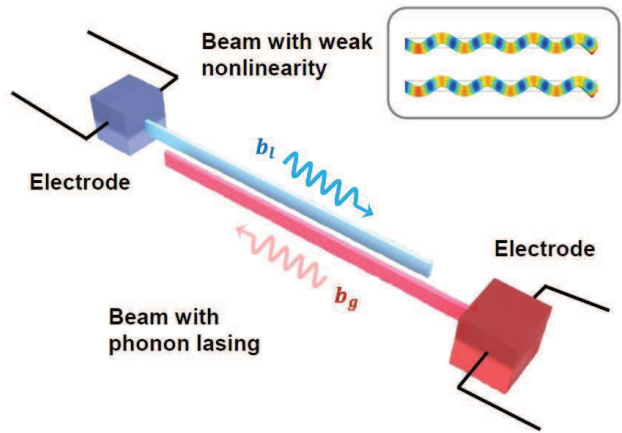}}\caption{(Color online) Schematic diagram of the
phonon diode system with two mechanical beams in which a beam with
weak mechanical nonlinearity is electrically or elastically
coupled to another beam with mechanical gain. The insets show the
finite-element-method (FEM) simulation by Comsol for the
mechanical modes.}\label{Fig of experimental setup}
\end{figure}

Let us now consider the design of the phonon diode system shown in
Fig.~\ref{Fig of experimental setup} in which a lossy vibrating
beam with damping rate $\Gamma_l$ and a weak Kerr
nonlinearity~\cite{RAlmogPRL:2007} of strength $\mu$ is
elastically coupled to another vibrating beam with gain
$\Gamma_g$~\cite{IMahboobPRL:2013}. The frequencies of the two
beams are both $\Omega_0$ and the mechanical coupling strength is
$g_{mm}$.

\begin{figure*}[t]\centerline{\includegraphics[width=16
cm]{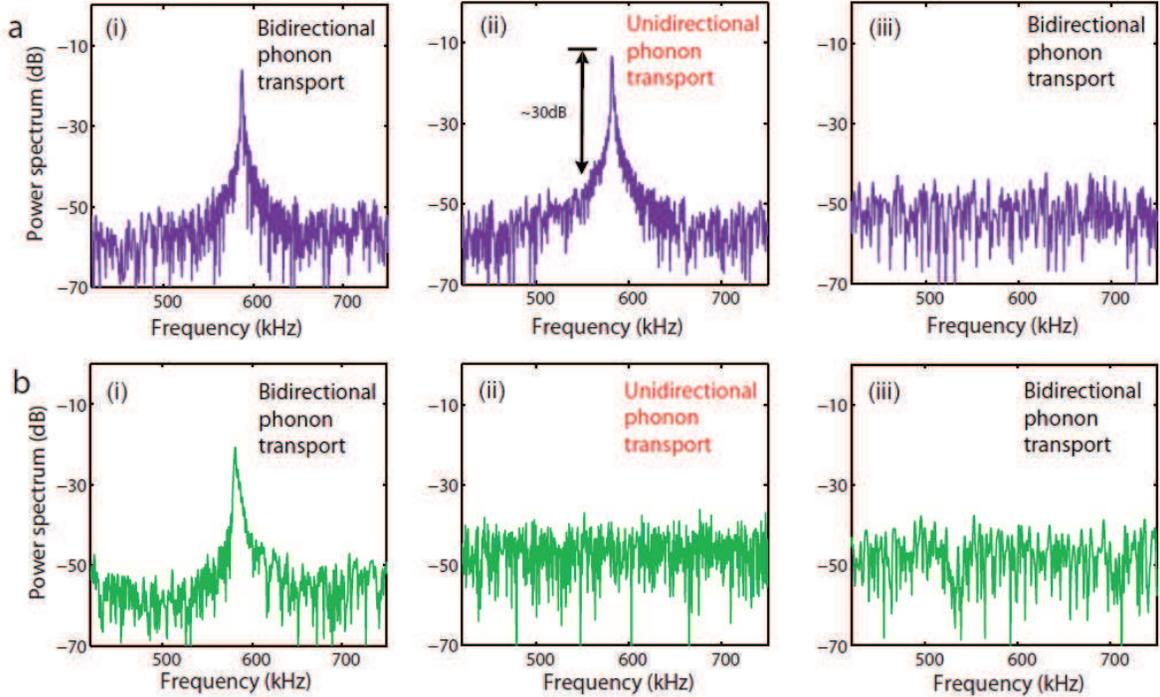}}\caption{(Color online) Numerical results
demonstrating unidirectional phonon transport in a
$\mathcal{PT}$-symmetric mechanical system in the
broken-$\mathcal{PT}$ phase. (a) Power spectrum obtained at the
output of the active beam without mechanical nonlinearity when the
phonon excitation (input) is at the passive beam with weak
nonlinearity. (b) Power spectrum obtained at the output of the
passive beam when the phonon excitation (input) is at the active
beam. When the intensity of the phonon excitation is within the
region bounded by Eq.~(\ref{Nonreciprocal window for input
intensity}), phonon transport is unidirectional. Transport from
the passive to the active resonator is allowed [see a(ii)], but
the transport from the active to the passive resonator is
prevented [b(ii)]. The rectification is about $30$ dB. If the
intensity of the phonon excitation is larger than the upper bound
of the unidirectional transport region, phonon transport is
bidirectional [a(i) and b(i)]. Phonon transport is not allowed in
either of the directions [a(iii) and b(iii)] if the intensity of
the phonon excitation is smaller than the lower bound of the
region given in Eq.~(\ref{Nonreciprocal window for input
intensity}).}\label{Fig of the power spectrum}
\end{figure*}

In Fig.~\ref{Fig of the power spectrum}, we present the numerical
results performed with the system parameters: $\Omega_0=600$ kHz,
$\Gamma_l=33$ kHz, $\Gamma_g=30$ kHz, $\delta=1.65$ kHz, $\mu=5.7$
kHz, and $g_{mm}=1$ kHz. Here, we fix the detuning $\delta$ and
change the amplitude of the input field. There is a $50$ dB
background noise which includes the combined effect of the thermal
noise on the mechanical resonators, the electrical noises induced
by the measurement apparatus and other possible sources of noise.
The results shown in Fig.~\ref{Fig of the power spectrum} for the
phonon diode agree well with the general model discussed in the
previous section. When the amplitude of the input field is
increased, it is clearly seen that there is a nonreciprocal region
in which phonon transport from the active beam to the passive beam
is almost completely suppressed (see Fig.~\ref{Fig of the power
spectrum}b(ii)), but phonon transport from the passive beam to the
active beam is allowed (see Fig.~\ref{Fig of the power
spectrum}a(ii)). A rectification ratio of about $30$ dB is
obtained. When the amplitude of the phonon excitation is larger
than the upper bound of the unidirectional phonon transport
region, the transport is bidirectional. In this case, the phonons
can freely move from the active beam to the passive beam and vice
versa (see Figs.~\ref{Fig of the power spectrum}a(i) and \ref{Fig
of the power spectrum}b(i)). Finally, for amplitudes of the phonon
excitation smaller than the lower bound of the region, no phonon
transport can take place between the resonators (see
Figs.~\ref{Fig of the power spectrum}a(iii) and \ref{Fig of the
power spectrum}b(iii)). These are the result of hysteresis (see
Fig.~\ref{Fig of unidirectional phonon transport}b) caused by the
strong mechanical nonlinearity.

\section{Discussions}\label{s6}
We have proposed a method to generate ultra-strong mechanical
nonlinearity with a very low-loss rate using a
$\mathcal{PT}$-symmetric mechanical structure in which a
mechanical resonator with gain but no nonlinearity is coupled to a
lossy (i.e., passive mechanical loss and no gain) mechanical
resonator with very weak nonlinearity. We have showed that the
weak mechanical nonlinearity is redistributed in the supermodes of
the coupled mechanical system and is enhanced (by more than three
orders of magnitude) when the mechanical $\mathcal{PT}$ system
enters the broken-$\mathcal{PT}$ regime. Moreover, owing to the
presence of the mechanical gain in one of the resonators to
compensate the mechanical loss of the other resonator, the
effective mechanical damping rate is decreased in the
$\mathcal{PT}$-symmetric system. Using experimentally accessible
parameter values, we identified the regimes where unidirectional
phonon transport is possible from the passive to active resonator
but not in the opposite direction. We then proposed an
experimentally-realizable system where a mechanical beam with
passive loss and weak nonlinearity is coupled to another beam
which acts like an active mechanical resonator. A possible
bottleneck for this design to achieve a phonon diode operated in
ambient condition is whether the mechanical gain observed with the
mechanical beams in a controlled environment and at low
temperatures~\cite{MHMathenyPRL:2014} could also be obtained in
ambient-temperature conditions. A possible way to overcome this
problem, and to realize phonon diodes in ambient conditions, is to
use a hybrid system composed of a gain optomechanical resonator
and an nonlinear electrically-driven mechanical
beam~\cite{RAlmogPRL:2007}, where the coupling between them is
achieved via the evanescent optical field of the optomechanical
resonator~\cite{GAnetsbergerNatPhys:2009}. The mechanical gain of
the optomechanical resonators can be provided at ambient
conditions by, e.g., the optomechanical dynamical instability in
the blue detuning regime~\cite{MarquardtPRL:2006}, which has been
demonstrated in optomechanical resonators in various
experiments~\cite{HRokhsariOE:2005}. Since creating
strongly-nonlinear mechanical or acoustic materials remains
challenging, we believe that the proposed system and the developed
approach provide a suitable platform for investigating nonlinear
phononics and can be used as a building block to design more
complex hybrid optomechanical or electromechanical information
processors. We envision that $\mathcal{PT}$ mechanical systems
will open a new route for designing functional phononic systems
with nonreciprocal phonon responses.
\\[0.2cm]

\begin{center}
\textbf{ACKNOWLEDGMENTS}
\end{center}

JZ is supported by the NSFC under Grant Nos. 61174084, 61134008.
YXL is supported by the NSFC under Grant Nos. 10975080, 61025022,
91321208. YXL and JZ are supported by the National Basic Research
Program of China (973 Program) under Grant No. 2014CB921401, the
Tsinghua University Initiative Scientific Research Program, and
the Tsinghua National Laboratory for Information Science and
Technology (TNList) Cross-discipline Foundation. LY and SKO are
supported by ARO grant No. W911NF-12-1-0026 and the NSFC under
Grant No. 61328502. F.N. is supported by the RIKEN iTHES Project,
MURI Center for Dynamic Magneto-Optics via the AFOSR award number
FA9550-14-1-0040, and Grant-in-Aid for Scientific Research (A).

\begin{center}
\textbf{AUTHOR CONTRIBUTIONS}
\end{center}

JZ, BP, SKO contributed equally to this work. LY, FN, SKO, YXL
supervised the project.
\\[0.2cm]

\appendix
\section{Nonlinearity enhancement by broken $\mathcal{PT}$ symmetry}\label{as1}

In order to prove the enhancement of mechanical nonlinearity in
the broken-$\mathcal{PT}$-symmetric regime, let us consider a
system of two coupled mechanical resonators, in which one of the
resonators has mechanical gain (active resonator) and thus a
positive damping rate $\Gamma_g$ and the second mechanical
resonator has a passive mechanical loss (passive resonator) with
loss rate $\Gamma_l$. The resonators have the same mechanical
frequency $\Omega_0$, and the annihilation operators for their
mechanical modes are denoted as $b_g$ and $b_l$, respectively, for
the active and passive resonators. Moreover, the passive
mechanical resonator has a weak mechanical Kerr-nonlinearity
denoted by $\mu$. The Hamiltonian describing these coupled
mechanical resonators can be written as
\begin{eqnarray}\label{Two coupled mechanical resonators}
H&=&\left(\Omega_0-i\Gamma_l\right)b_l^{\dagger}b_l+\left(\Omega_0+i\Gamma_g\right)b_g^{\dagger}b_g\nonumber\\
&&+g_{mm}
\left(b_l^{\dagger}b_g+b_lb_g^{\dagger}\right)+\mu\left(b_l^{\dagger}b_l\right)^2,
\end{eqnarray}
where $g_{mm}$ is the coupling strength between the mechanical
modes of the resonators. Generally, the nonlinear Kerr term in
Eq.~(\ref{Two coupled mechanical resonators}) will shift the
boundary between the $\mathcal{PT}$ symmetric regime and the
broken-$\mathcal{PT}$ regime. However, in our model, the Kerr
nonlinearity denoted by $\mu$ is very weak, and we can omit the
nonlinearity-induced shift of this boundary. To find the boundary
of $\mathcal{PT}$ transition, we consider the first three terms in
Eq.~(\ref{Two coupled mechanical resonators})
\begin{eqnarray}\label{Linear part of the PT system}
H_1&=&\left(\Omega_0-i\Gamma_l\right)b_l^{\dagger}b_l+\left(\Omega_0+i\Gamma_g\right)b_g^{\dagger}b_g\nonumber\\
&&+g_{mm} \left(b_l^{\dagger}b_g+b_lb_g^{\dagger}\right),
\end{eqnarray}
which can be written as
\begin{equation}\label{Matrix form of the linear Hamiltonian of the PT system}
H_1=\left(%
\begin{array}{cc}
  b_g^{\dagger} & b_l^{\dagger} \\
\end{array}%
\right)\left(%
\begin{array}{cc}
  \Omega_0+i\Gamma_g & g_{mm} \\
  g_{mm} & \Omega_0-i\Gamma_l \\
\end{array}%
\right)\left(%
\begin{array}{c}
  b_g \\
  b_l \\
\end{array}%
\right).
\end{equation}
This Hamiltonian can be diagonalized as
\begin{equation}\label{Diagonalized PT Hamiltonian}
H_1=\left(%
\begin{array}{cc}
  b_g^{\dagger} & b_l^{\dagger} \\
\end{array}%
\right)P^{-1}\left(%
\begin{array}{cc}
  \Omega_+-i\Gamma_+ & 0 \\
  0 & \Omega_--i\Gamma_- \\
\end{array}%
\right)P\left(%
\begin{array}{c}
  b_g \\
  b_l \\
\end{array}%
\right),
\end{equation}
where the transformation matrix $P$ is defined by
\begin{equation}\label{Transformation matrix}
P=\frac{\left(%
\begin{array}{cc}
  g_{mm} & \left[\left(\Omega_+-\Omega_0\right)-i\left(\Gamma_+-\Gamma_l\right)\right] \\
  g_{mm} & \left[\left(\Omega_--\Omega_0\right)-i\left(\Gamma_--\Gamma_l\right)\right] \\
\end{array}%
\right)}{\sqrt{\left(\Omega_{\pm}-\Omega_0\right)^2+\left(\Gamma_{\pm}-\Gamma_l\right)^2+g_{mm}^2}}.
\end{equation}
Consequently, we have
\begin{equation}\label{Supermodes}
\left(%
\begin{array}{c}
  b_+ \\
  b_- \\
\end{array}%
\right)=P\left(%
\begin{array}{c}
  b_g \\
  b_l \\
\end{array}%
\right)
\end{equation}
as the mechanical supermodes formed by the coupling of the
resonators. These supermodes $b_{\pm}$ are characterized by the
eigenfrequencies $\Omega_{\pm}$ and damping rates $\Gamma_{\pm}$.
\begin{figure} \centerline{\includegraphics[width =
7.6 cm]{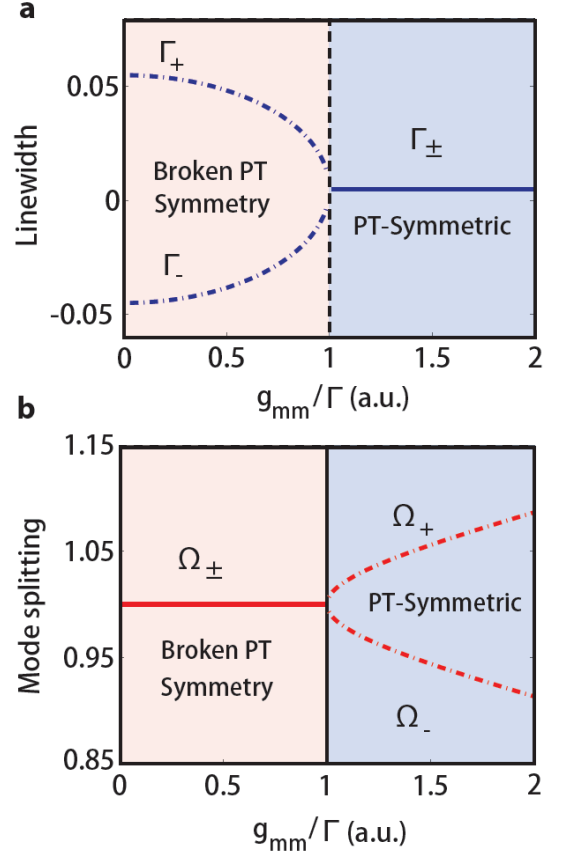}}\caption{(Color online). Evolution of the
eigenfrequencies of the coupled mechanical resonators. (a)
Difference of the real parts of the eigenfrequencies of the
supermodes: mode splitting, and (b) difference of the imaginary
parts of the eigenfrequencies (i.e., linewidth) of the supermodes.
The resonance frequencies of the supermodes are non-degenerate in
the $\mathcal{PT}$-symmetric regime. In the broken-$\mathcal{PT}$
symmetry regime, however, they are frequency
degenerate.}\label{Fig of eigenfrequencies of the PT system}
\end{figure}

For this mechanical $\mathcal{PT}$ symmetric system, there are two
different regimes (see Fig.~\ref{Fig of eigenfrequencies of the PT
system}):

\noindent (i) $\mathcal{PT}$ symmetric regime where
\begin{equation}\label{Condition for PT symmetry}
\Gamma=\frac{\left(\Gamma_l+\Gamma_g\right)}{2}\leq g_{mm},
\end{equation}
and the two supermodes $b_+$ and $b_-$ are nondegenerate in their
resonance frequencies (i.e., real part of their complex
eigenfrequencies) given by
\begin{equation}\label{Frequency in the PT symmetric regime}
\Omega_{\pm}=\Omega_0\pm\beta=\Omega_0\pm\sqrt{g_{mm}^2-\Gamma^2}.
\end{equation}
The damping rates of the supermodes (i.e., linewidths of the
resonances; imaginary part of their complex eigenfrequencies) are
the same and equal to
\begin{equation}\label{Damping rate in the PT symmetric regime}
\Gamma_{\pm}=\chi=\frac{\Gamma_l-\Gamma_g}{2}.
\end{equation}

\noindent (ii) Broken $\mathcal{PT}$-symmetry regime where
\begin{equation}\label{Condition for PT symmetric breaking}
\Gamma=\frac{\Gamma_l+\Gamma_g}{2}>g_{mm}.
\end{equation}
The two supermodes $b_+$ and $b_-$ are degenerate in their
resonance frequencies
\begin{equation}\label{Frenquency in the broken-PT regime}
\Omega_{\pm}=\Omega_0,
\end{equation}
and their damping rates are different:
\begin{equation}\label{Damping rate in the broken-PT regime}
\Gamma_{\pm}=\chi\mp i\beta.
\end{equation}
Now let us consider the nonlinear Kerr term in Eq.~(\ref{Two
coupled mechanical resonators}). Using Eq.~(\ref{Supermodes}), we
find
\begin{eqnarray*}
b_l&=&\frac{\sqrt{\left(\Omega_+-\Omega_0\right)^2+\left(\Gamma_+-\Gamma_l\right)^2+g_{mm}^2}}{\left(\Omega_+-\Omega_-\right)-i\left(\Gamma_+-\Gamma_-\right)}b_+\\
&&-\frac{\sqrt{\left(\Omega_--\Omega_0\right)^2+\left(\Gamma_--\Gamma_l\right)^2+g_{mm}^2}}{\left(\Omega_+-\Omega_-\right)-i\left(\Gamma_+-\Gamma_-\right)}b_-\\
&=&\beta_{l+}b_++\beta_{l-}b_-.
\end{eqnarray*}
By substituting the above equation into the last term on the right
hand side of Eq.~(\ref{Two coupled mechanical resonators}) and
dropping the non-resonant terms, we can rewrite the nonlinear Kerr
term of Eq.~(\ref{Two coupled mechanical resonators}) as
\begin{equation}\label{Nonlinear terms}
H_{nl}=\left(|\beta_{l+}|^2b_+^{\dagger}b_++|\beta_{l-}|^2b_-^{\dagger}b_-\right
)^2.
\end{equation}
The self-Kerr terms
$|\beta_{l+}|^4\left(b_+^{\dagger}b_+\right)^2$ and
$|\beta_{l-}|^4\left(b_-^{\dagger}b_-\right)^2$ only lead to a
frequency-shift of the two supermodes and thus are less important.
The cross-Kerr term
\begin{equation}\label{The cross-Kerr term}
H_{nl}^{\prime}=|\beta_{l+}|^2|\beta_{l-}|^2\left(b_+^{\dagger}b_+\right)\left(b_-^{\dagger}b_-\right)=\mu^{\prime}\left(b_+^{\dagger}b_+\right)\left(b_-^{\dagger}b_-\right)
\end{equation}
is more important and leads to the redistribution of the nonlinear
effect among the two supermodes. From Eqs.~(\ref{Condition for PT
symmetry})-(\ref{Damping rate in the PT symmetric regime}), the
nonlinear coefficient $2|\beta_{l+}|^2|\beta_{l-}|^2$ can be
represented in the broken-$\mathcal{PT}$ regime as $\mu_b'$, and
in the $\mathcal{PT}$ symmetric regime as $\mu_s'$
\begin{equation}\label{Nonlinear coefficients in different regimes}
\mu^{\prime}_b=\mu\frac{\Gamma^2g_{mm}^2}{\left(\Gamma^2-g_{mm}^2\right)^2},\quad
\mu^{\prime}_s=\mu\frac{g_{mm}^4}{\left(\Gamma^2-g_{mm}^2\right)^2}.
\end{equation}

As was observed in photonic
experiments~\cite{BPengNatPhys2014,BPengScience:2014}, in the
broken-$\mathcal{PT}$ regime the two supermodes $b_{\pm}$ are
degenerate and the field is localized  in the gain resonator, and
thus the field $b_l$ is much smaller than $b_g$. Therefore, we can
omit the terms related to $b_l$ in the expressions of the
supermodes $b_{\pm}$ and we have
\begin{eqnarray*}
b_+\approx
\frac{g_{mm}}{\sqrt{\left(\Omega_+-\Omega_0\right)^2+\left(\Gamma_+-\Gamma_l\right)^2+g_{mm}^2}}b_g,\\
b_-\approx
\frac{g_{mm}}{\sqrt{\left(\Omega_--\Omega_0\right)^2+\left(\Gamma_--\Gamma_l\right)^2+g_{mm}^2}}b_g.
\end{eqnarray*}
Subsequently, we find that the cross-Kerr term given in
Eq.~(\ref{The cross-Kerr term}) can induce a self-Kerr effect in
the gain resonator
\begin{equation}\label{Self-Kerr term in the PT breaking regime}
H_{nl}^{\prime}=\mu\frac{g_{mm}^4}{4\left(\Gamma^2-g_{mm}^2\right)^2}\left(b_g^{\dagger}b_g\right)^2.
\end{equation}
Clearly, when $\Gamma\approx  g_{mm}$ (in the vicinity of the
spontaneous $\mathcal{PT}$-symmetry breaking point: the
$\mathcal{PT}$-phase transition point), this self-Kerr
nonlinearity is greatly enhanced.

\section{Unidirectional phonon transport by mechanical
nonlinearity}\label{as2} Let us now present a detailed analysis
for finding the unidirectional phonon transport region near the
$\mathcal{PT}$-transition point. In this case, the gain-loss
balance between the active resonator, with annihilation operator
$b_g$, and the passive resonator, with annihilation operator
$b_l$, decreases the effective damping rates of the two modes. In
the vicinity of the $\mathcal{PT}$-phase transition point (i.e.,
$\Gamma\approx g_{mm}$), the effective damping rates of the two
modes is given by $\chi=\left(\Gamma_l-\Gamma_g\right)/2$. The
coupling between the two mechanical resonators also leads to the
transfer of mechanical Kerr nonlinearity from the passive
resonator to the active resonator, and this mechanical
nonlinearity is strongly enhanced near the
$\mathcal{PT}$-transition point (i.e., $\Gamma\approx g_{mm}$).
Hereafter, we will denote this enhanced mechanical Kerr
nonlinearity coefficient as $\mu_b'$.

Let us first consider the phonon transport from the passive
resonator to the active resonator. Here the phononic field in the
passive resonator is generated via an phononic input field with
strength $\varepsilon_d$ and frequency $\Omega_d$. Using the
standard input-output
formalism~\cite{CWGardinerBook,CWGardinerPRA}, the output field of
the active mechanical resonator is found as $b_{\rm
out}=\chi^{1/2}b_g$, which shows that the output field is
proportional to the intracavity field $b_g$, if we omit the vacuum
fluctuations in the input field. Thus the transmission from
passive to active resonator is given by
\begin{equation}\label{Power transmittance from loss to gain resonator in appendix}
T_{l\rightarrow g}\left(\delta\right)=\chi
n_g/\left|\varepsilon_d\right|^2,
\end{equation}
where $n_g$
represents the steady-state value of the intracavity phonon number
in the active resonator. From the steady-state solution of the
equations of motion for the coupled mechanical resonator system,
we find that $n_g$ satisfies
\begin{equation}\label{Stationary equation of ng}
\tilde{\mu}^2n_g^3-2\tilde{\mu}\tilde{\Omega}n_g^2+\left(\tilde{\Gamma}^2+\tilde{\Omega}^2\right)n_g-\tilde{n}_{\rm
in}=0,
\end{equation}
where
\begin{eqnarray*}
&\tilde{\Gamma}=\left(\chi^2+\delta^2+g_{mm}^2\right)\chi,\quad
\tilde{\Omega}=\left(\chi^2+\delta^2\right)\Omega_0-g_{mm}
^2\delta,&\\
&\tilde{\mu}=\left(\chi^2+\delta^2\right)\mu_b^{\prime},\quad
\tilde{n}_{\rm
in}=|\varepsilon_d|^2g_{mm}^2\left(\chi^2+\delta^2\right).&
\end{eqnarray*}
The algebraic equation~(\ref{Stationary equation of ng}) has three
or one root depending on the system parameters, and one of the
roots is unstable if the algebraic equation~(\ref{Stationary
equation of ng}) has three roots. When we increase the detuning
$\delta=\Omega_0-\Omega_d$, such that
\begin{equation}\label{Bistable condition}
\frac{\Omega_0-\left(g_{mm}^2\delta\right)/\left(\chi^2+\delta^2\right)}{\chi\left(\chi^2+\delta^2+g_{mm}^2\right)/\left(\chi^2+\delta^2\right)}=\sqrt{3},
\end{equation}
or equivalently,
\begin{widetext}
\begin{equation}\label{Equivalent bistable condition}
\delta=\delta_{\rm
max}=\frac{g_{mm}^2+\sqrt{g_{mm}^4+4\left(\sqrt{3}\chi^3-\chi^2\Omega_0+\sqrt{3}g_{mm}^2\chi\right)\left(\Omega_0-\sqrt{3}\chi\right)}}{2\left(\Omega_0-\sqrt{3}\chi\right)},
\end{equation}
\end{widetext}
the system enters the bistable regime. In fact, when
$\delta\leq\delta_{\rm max}$, the algebraic equation has three
branches of solutions. However, two branches of solutions
disappear when $\delta>\delta_{\rm min}$ (see
Ref.~\cite{IVBarashenkovPRE:1996} and the supplementary materials
of Ref.~\cite{THerr}). In this case, the transmittance of the
photon transport $T_{l\rightarrow g}\left(\delta\right)$ changes
suddenly from a high value to a low value. Noting that
$g_{mm}\gg\chi$ near the $\mathcal{PT}$ breaking point, the
critical detuning $\delta_{\rm max}$ can be approximately
estimated to be
\begin{equation}\label{Upper bound of the unidirectional phonon transport window}
\delta_{\rm max}=\frac{g_{mm}^2}{\Omega_0-\sqrt{3}\chi}.
\end{equation}

Let us now consider the phonon transport from the active
mechanical resonator to the passive one. The driving field with
strength $\varepsilon_d$ and frequency $\Omega_d$ is fed into the
gain resonator in this case. Following the same discussion and
approach as for the previous case, it can be shown that a
bistability-induced phase transition occurs when the detuning
$\delta=\Omega_0-\Omega_d$ satisfies
\begin{equation}\label{Bistable condition from gain to loss}
\frac{\delta-\left(g_{mm}^2\Omega_0\right)/\left(\chi^2+\Omega_0^2\right)}{\chi\left(\chi^2+\Omega_0^2+g_{mm}^2\right)/\left(\chi^2+\Omega_0^2\right)}=\sqrt{3},
\end{equation}
or equivalently,
\begin{equation}\label{Equivalent bistable condition from gain to loss}
\delta=\delta_{\rm
min}=\frac{\sqrt{3}\left(\chi^2+\Omega_0^2+g_{mm}^2\right)\chi+g_{mm}^2\Omega_0}{\chi^2+\Omega_0^2}.
\end{equation}
Near the $\mathcal{PT}$-breaking point, $\chi\ll
g_{mm},\,\Omega_0$, and thus $\delta_{\rm min}$ can be
approximately estimated to be
\begin{equation}\label{Lower bound of the unidirectional phonon transport window}
\delta_{\rm min}=\frac{g_{mm}^2\Omega_0}{\chi^2+\Omega_0^2}.
\end{equation}
Combing Eqs.~(\ref{Upper bound of the unidirectional phonon
transport window}) and (\ref{Lower bound of the unidirectional
phonon transport window}), we find that when the detuning $\delta$
is within the following region
\begin{equation}\label{Unidirectional phonon transport
window} \left[\delta_{\rm min},\delta_{\rm
max}\right]=\left[\frac{g_{mm}^2\Omega_0}{\Omega_0^2+\chi^2},\frac{g_{mm}^2}{\Omega_0-\sqrt{3}\chi}\right],
\end{equation}
it is possible to observe the unidirectional phonon transport,
i.e., the phonon transport from the passive resonator to the
active resonator is allowed, whereas the phonon transport from the
active resonator to the passive resonator is blocked.

Figure.~\ref{Fig of unidirectional phonon transport window}a shows
the transmittance functions $T_{l\rightarrow
g}\left(\delta\right)$ and $T_{g\rightarrow l}\left(\delta\right)$
as a function of the detuning $\delta$. It is (as explained in the
main text) clear that there is a unidirectional phonon transport
region when the detuning is up-scanned from smaller to larger
detuning. We also show in Fig.~\ref{Fig of unidirectional phonon
transport window}b the rectification ratios for up-scanning and
down-scanning the detuning $\delta$. Similar to our previous
discussions, a non-reciprocal region can be observed for the
up-scanning process, while it disappears for the down-scanning
process, and a high rectification-ratio, larger than $30$ dB, can
be obtained within the nonreciprocal region.
\begin{figure*}[t] \centerline{\includegraphics[width =
16.0 cm]{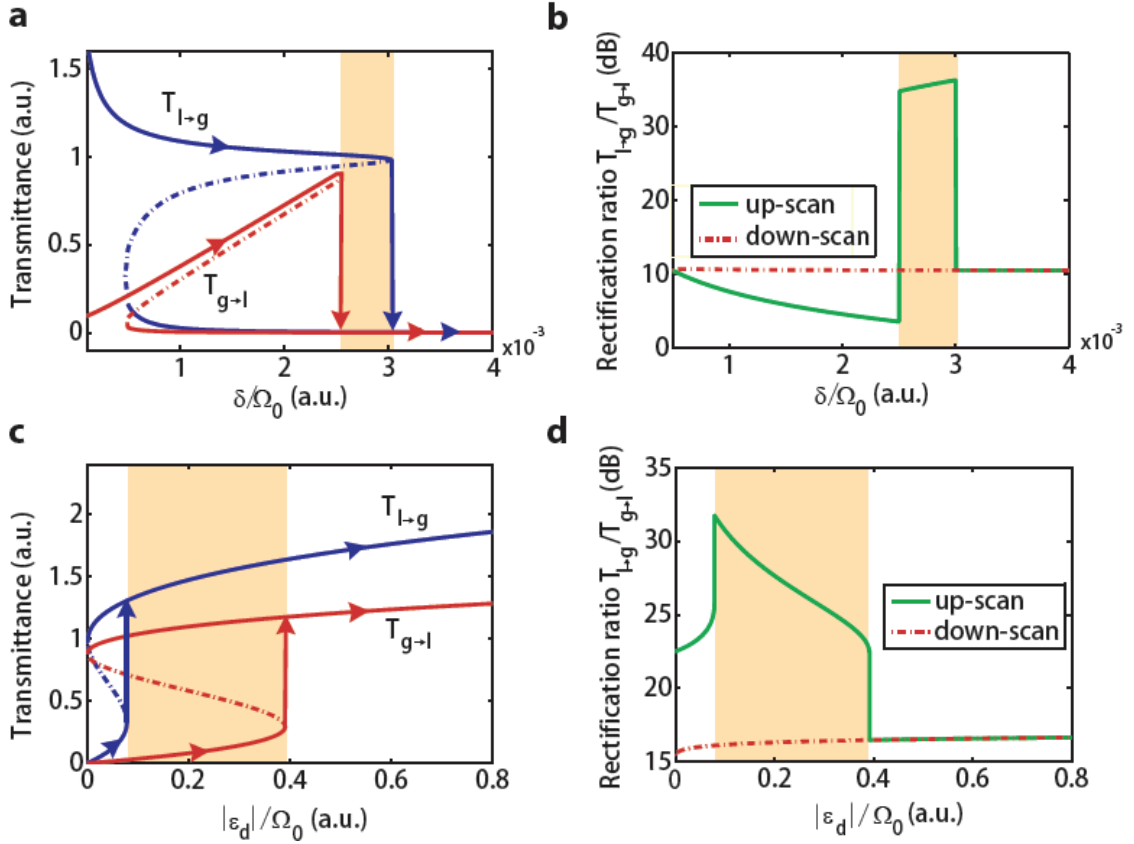}}\caption{(Color online). Bistability curves
and unidirectional phonon-transport regions. (a) Transmittances as
a function of the detuning frequency $\delta$, when the input
field amplitude is fixed at $\varepsilon_d$. (b) Rectification
ratio for the bidirectional phonon transport versus the detuning
$\delta$: a rectification ratio larger than $30$ dB can be
obtained when the detuning is up-scanned to enter the
unidirectional phonon transport region. (c) Transmittances as
function of the intensity of the input field when the detuning
frequency $\delta$ is fixed and its value is taken within the
unidirectional phonon transport region in (a). The blue and red
curves represent the power transmittances from the passive to the
active resonator $T_{l\rightarrow g}$ and from the active to the
passive resonator $T_{g\rightarrow l}$. The solid and dashed parts
on each curve denote the stable and unstable solutions of the
bistable system. The unstable solutions cannot be observed in the
output and thus lead to sudden transitions (black solid arrows) in
the transmittance functions. (d) Rectification ratios versus
normalized amplitude of the input field for fixed detuning
$\delta$. The melon-colored shaded areas denote the unidirectional
transport regions.}\label{Fig of unidirectional phonon transport
window}
\end{figure*}

Up to this point, we do not consider the amplitude of the input
field. Let us assume that the detuning $\delta$ is fixed and is
within the detuning region given by Eq.~(\ref{Unidirectional
phonon transport window}). We then vary the amplitude of the input
field to show the bistability and the hysteresis in the
transmittance functions. Let us first assume that $\delta>g_{mm}$.
If we consider the phonon transport from the passive resonator to
the active resonator, we can obtain an algebraic equation similar
to that given in Eq.~(\ref{Stationary equation of ng}). The
bistable transition point corresponds to the stationary points of
the function
\begin{equation}\label{function for the upper bound of the input intensity window}
f(n_g)=\tilde{\mu}^2n_g^3-2\tilde{\mu}\tilde{\Omega}n_g^2+\left(\tilde{\Gamma}^2+\tilde{\Omega}^2\right)n_g.
\end{equation}
By setting $f'(n_g)=0$, the stationary point of $f(n_g)$ can be
found as
\begin{equation}\label{ng for maximum input intensity}
n_g^*=\left[2\tilde{\mu}\tilde{\Omega}-\sqrt{4\tilde{\mu}^2\tilde{\Omega}^2-3\tilde{\mu}^2\left(\tilde{\Gamma}^2+\tilde{\Omega}^2\right)}\right]\left(3\tilde{\mu}^2\right)^{-1}.
\end{equation}
The upper bound of the unidirectional phonon transport region is
given by
\begin{widetext}
\begin{eqnarray*}
\left|\varepsilon_{\rm
max}\right|^2=\frac{f(n_g^*)}{g_{mm}^2\left(\chi^2+\delta^2\right)}=\frac{2\tilde{\Omega}\left(\tilde{\Gamma}^2+\tilde{\Omega}^2\right)}{9\tilde{\mu}g_{mm}^2\left(\chi^2+\delta^2\right)}+\frac{\left(6\tilde{\Gamma}^2-2\tilde{\Omega}^2\right)}{9g_{mm}^2\left(\chi^2+\delta^2\right)}\left[\frac{2\tilde{\mu}\tilde{\Omega}-\sqrt{4\tilde{\mu}^2\tilde{\Omega}^2-3\tilde{\mu}^2\left(\tilde{\Gamma}^2+\tilde{\Omega}^2\right)}}{3\tilde{\mu}^2}\right].
\end{eqnarray*}
\end{widetext}
Near the $\mathcal{PT}$-transition point, we have
$\delta,\,g_{mm}\gg\chi$, and thus it can be approximately
estimated that
\begin{equation}\label{Upper bound of the power intensity nonreciprocal window}
\left|\varepsilon_{\rm
max}\right|^2\approx\frac{2\left(\delta^2+g_{mm}^2\right)^2}{9\mu_b'
g_{mm}^2\delta}.
\end{equation}
Let us now consider the case of phonon transport from the active
resonator to the passive resonator when the amplitude of the input
field is varied and the detuning is kept fixed. In this case, we
obtain
\begin{equation}\label{algebraic equation for the lower bound for input intensity}
\tilde{\tilde{\mu}}^2
n_g^3-2\tilde{\tilde{\mu}}\tilde{\tilde{\Omega}}n_g^2+\left(\tilde{\tilde{\Gamma}}^2+\tilde{\tilde{\Omega}}^2\right)n_g-\tilde{\tilde{n}}_{\rm
in}=0,
\end{equation}
where
\begin{eqnarray*}
&\tilde{\tilde{\Gamma}}=\left(\chi^2+\Omega_0^2+g_{mm}^2\right)\chi,\quad\tilde{\tilde{\Omega}}=\left(\chi^2+\Omega^2\right)\delta-g_{mm}^2\Omega_0,&\\
&\tilde{\tilde{\mu}}=\left(\chi^2+\Omega_0^2\right)\mu_b',\quad\tilde{\tilde{n}}_{\rm
in}=\left(\chi^2+\delta^2\right)^2\left|\varepsilon_d\right|^2.&
\end{eqnarray*}
Similar to Eq.~(\ref{function for the upper bound of the input
intensity window}), the bistable transition point can be found by
calculating the stationary points of the function
\begin{equation}\label{function for the lower bound of the input intensity window}
f(n_g)=\tilde{\tilde{\mu}}^2
n_g^3-2\tilde{\tilde{\mu}}\tilde{\tilde{\Omega}}n_g^2+\left(\tilde{\tilde{\Gamma}}^2+\tilde{\tilde{\Omega}}^2\right)n_g.
\end{equation}
which leads to
\begin{equation}\label{ng for minimum input intensity}
\tilde{n}_g^*=\left[2\tilde{\tilde{\mu}}\tilde{\tilde{\Omega}}-\sqrt{4\tilde{\tilde{\mu}}^2\tilde{\tilde{\Omega}}^2-3\tilde{\tilde{\mu}}^2\left(\tilde{\tilde{\Gamma}}^2+\tilde{\tilde{\Omega}}^2\right)}\right]\left(3\tilde{\tilde{\mu}}^2\right)^{-1}.
\end{equation}
The lower bound of the unidirectional phonon transport region is
then given by
\begin{eqnarray*} \left|\varepsilon_{\rm
min}\right|^2&=&\frac{\tilde{f}(\tilde{n}_g^*)}{\left(\chi^2+\delta^2\right)^2}=\frac{2\tilde{\tilde{\Omega}}\left(\tilde{\tilde{\Gamma}}^2+\tilde{\tilde{\Omega}}^2\right)}{9\tilde{\tilde{\mu}}\left(\chi^2+\delta^2\right)^2}+\frac{\left(6\tilde{\tilde{\Gamma}}^2-2\tilde{\tilde{\Omega}}^2\right)}{9g_{mm}^2\left(\chi^2+\delta^2\right)}\\
&&\left[\frac{2\tilde{\tilde{\mu}}\tilde{\tilde{\Omega}}-\sqrt{4\tilde{\tilde{\mu}}^2\tilde{\tilde{\Omega}}^2-3\tilde{\tilde{\mu}}^2\left(\tilde{\tilde{\Gamma}}^2+\tilde{\tilde{\Omega}}^2\right)}}{3\tilde{\tilde{\mu}}^2}\right].
\end{eqnarray*}
Near the $\mathcal{PT}$-transition point, we have
$\delta,\,g_{mm}\gg\chi$, and it can be approximately estimated
that
\begin{equation}\label{Lower bound of the power intensity nonreciprocal window}
\left|\varepsilon_{\rm
min}\right|^2\approx\frac{2\left(\delta^2+g_{mm}^2\right)^2}{9\mu_b'
\delta^3}.
\end{equation}
We thus conclude that nonreciprocal phonon transport takes place
if the amplitude of the input is within the region
\begin{equation}\label{Nonreciprocal window for input
intensity for strong coupling}
\left|\varepsilon_d\right|^2\in\left[\frac{2\left(\delta^2+g_{mm}^2\right)^3}{9\mu_b'\delta^3},\frac{2\left(\delta^2+g_{mm}^2\right)^3}{9\mu_b'g_{mm}^2\delta}\right].
\end{equation}
Similarly, when $\delta\leq g_{mm}$, the nonreciprocal region for
the amplitude of the input field can be written as
\begin{equation}\label{Nonreciprocal window for input
intensity for weak coupling}
\left|\varepsilon_d\right|^2\in\left[\frac{2\left(\delta^2+g_{mm}^2\right)^3}{9\mu_b'g_{mm}^2\delta},\frac{2\left(\delta^2+g_{mm}^2\right)^3}{9\mu_b'\delta^3}\right].
\end{equation}
In Fig.~\ref{Fig of unidirectional phonon transport window}c, we
present the transmittances as a function of the amplitude of the
input field when the detuning is kept fixed within the
unidirectional transport region given in Eq.~(\ref{Unidirectional
phonon transport window}). We see that the lower stable branches
of the bistable curves shown in Fig.~\ref{Fig of unidirectional
phonon transport window}c (the parts of the bistable curves before
the bistable transitions occur) increase when we increase the
intensity of the input field. This decreases the rectification, as
shown in Fig.~\ref{Fig of unidirectional phonon transport
window}d.

\section{Can this system be used as a phonon isolator?}\label{as3}
In order to check the performance of the proposed system as an
isolator for phonons, we study the system considering that phonons
are injected in the system in both directions, that is
simultaneously at the passive and active resonator sides. If the
system exhibits unidirectional phonon transport under this
condition, then the proposed system can be used as an isolator.
\begin{figure} \centerline{\includegraphics[width =
7.6 cm]{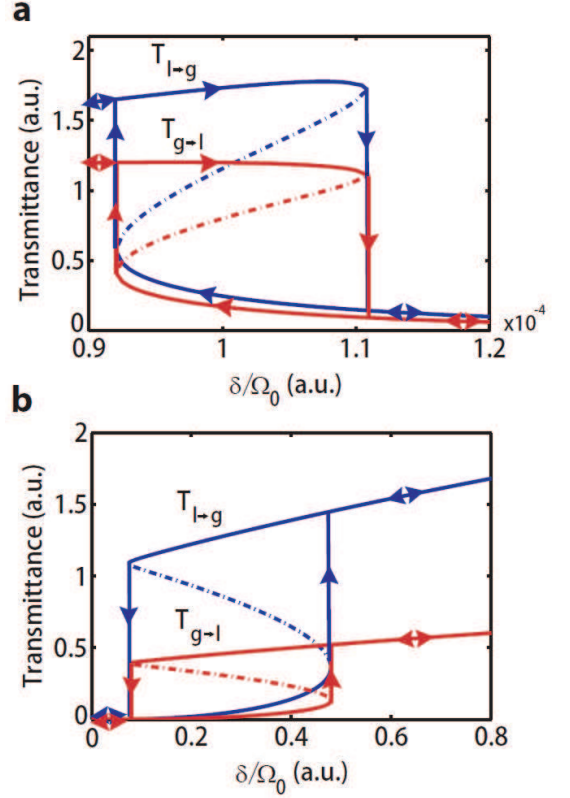}}\caption{(Color online). Bistability curves for
the mechanical $\mathcal{PT}$ system when phonons are input
simultaneously in both directions. (a) Transmittances as functions
of the detuning frequency $\delta$ when the input field amplitude
is fixed at $\varepsilon_d$. (b) Transmittances as functions of
the amplitude of the input field when the detuning frequency
$\delta$ is fixed. The blue and red curves represent the power
transmittance functions $T_{l\rightarrow g}$ and $T_{g\rightarrow
l}$. The solid and dashed parts on each curve denote the stable
and unstable solutions of the bistable systems. The unstable
solutions cannot be observed in the output and thus lead to sudden
transitions in the transmittance functions.}\label{Fig of bistable
curves for phonon isolator}
\end{figure}

The equations of motion of the system for this case can be written
as
\begin{eqnarray}\label{Dynamical equation for phonon isolator}
&\dot{b}_l=-\left(\chi+i\delta_l\right)b_l-i
g_{mm}b_g+i\varepsilon_l,&\nonumber\\
&\dot{b}_g=-\left(\chi+i\delta_g\right)b_g-i\mu_b'\left(b_g^{\dagger}b_g\right)b_g-ig_{mm}b_l+i\varepsilon_g,&\nonumber\\
\end{eqnarray}
where the last terms on the right-hand-sides of
Eq.~(\ref{Dynamical equation for phonon isolator}) denote the
input fields. The steady-state solution of Eq.~(\ref{Dynamical
equation for phonon isolator}) leads to
\begin{equation}\label{algebraic equation for phonon isolator}
\bar{\mu}^2n_g^3-2\bar{\mu}\bar{\Omega}n_g^2+\left(\bar{\Gamma}^2+\bar{\Omega}^2\right)n_g-\bar{n}_{\rm
in}=0,
\end{equation}
where
\begin{eqnarray*}
&\bar{\Gamma}=\left(\chi^2+\delta_l^2+g_{mm}^2\right)\chi,\quad \bar{\mu}=\left(\chi^2+\delta_l^2\right)\mu_b'&\\
&\bar{\Omega}=\left(\chi^2+\delta_l^2\right)\delta_g-g_{mm}^2\delta_l,&\\
&\bar{n}_{\rm
in}=|\varepsilon_l|^2g_{mm}^2\left(\chi^2+\delta_l^2\right)+\left(\chi^2+\delta_l^2\right)^2|\varepsilon_g|^2.&
\end{eqnarray*}
Let us first fix $\varepsilon_l,\,\varepsilon_g,\,\delta_g$, and
vary the detuning $\delta_l=\delta$. In this case, the bistable
transitions for both directions occur when the detuning $\delta$
satisfies
\begin{equation}\label{Bistable transition condition for phonon isolator}
\frac{\delta_g-\left(g_{mm}^2\delta\right)/\left(\chi^2+\delta^2\right)}{\chi\left(\chi^2+\delta^2+g_{mm}^2\right)/\left(\chi^2+\delta^2\right)}=\sqrt{3}.
\end{equation}
When the detuning is up-scanned from smaller to larger detuning
values, the bistable transition occurs for
\begin{eqnarray}\label{detuning from smaller to larger}
\delta&=&\sqrt{\frac{g_{mm}^4}{4\left(\delta_g-\sqrt{3}\chi\right)^2}+\frac{\left(\sqrt{3}\chi^3-\chi^2\delta_g+\sqrt{3}g_{mm}^2\chi\right)}{\left(\delta_g-\sqrt{3}\chi\right)}}\nonumber\\
&&+\frac{g_{mm}^2}{2\left(\delta_g-\sqrt{3}\chi\right)}.
\end{eqnarray}
When the detuning $\delta$ is down-scanned from larger to smaller
detuning values, the bistable transition occur at
\begin{eqnarray}\label{detuning from larger to smaller}
\delta&=&-\sqrt{\frac{g_{mm}^4}{4\left(\delta_g-\sqrt{3}\chi\right)^2}+\frac{\left(\sqrt{3}\chi^3-\chi^2\delta_g+\sqrt{3}g_{mm}^2\chi\right)}{\left(\delta_g-\sqrt{3}\chi\right)}}\nonumber\\
&&+\frac{g_{mm}^2}{2\left(\delta_g-\sqrt{3}\chi\right)}.
\end{eqnarray}
The transmittances presented in Fig.~\ref{Fig of bistable curves
for phonon isolator}a clearly show the bistable operation. A close
look at Fig.~\ref{Fig of bistable curves for phonon isolator}a
reveals that the transition from the bistable region to the stable
trajectories takes place at the same points for both directions.
We cannot find a detuning region within which transport in one
direction is allowed and the transport in the other direction is
prevented. Thus, we conclude that when phonons are injected
simultaneously at both input ports, we cannot see a unidirectional
operation. Consequently, it is impossible to use this system as an
isolator for phonons.

Let us now fix $\delta_l$, $\delta_g$, $\varepsilon_g$, and vary
$\varepsilon_l=\varepsilon_d$, to check the possibility of
providing a phonon isolator. The bistable transition point is just
the stationary points of the function
\begin{equation}\label{algebraic function to solve ng for phonon isolator}
\bar{f}\left(n_g\right)=\bar{\mu}^2
n_g^3-2\bar{\mu}\bar{\Omega}n_g^2+\left(\bar{\Gamma}^2+\bar{\Omega}^2\right)n_g-|\varepsilon_g|^2.
\end{equation}
By setting $\bar{f}'\left(n_g\right)=0$, we find
\begin{eqnarray*}
\bar{n}_{g1}^*&=&\left[2\bar{\mu}\bar{\Omega}+\sqrt{4\bar{\mu}^2\bar{\Omega}^2-3\bar{\mu}^2\left(\bar{\Gamma}^2+\bar{\Omega}^2\right)}\right]\left(3\bar{\mu}^2\right)^{-1},\\
\bar{n}_{g2}^*&=&\left[2\bar{\mu}\bar{\Omega}-\sqrt{4\bar{\mu}^2\bar{\Omega}^2-3\bar{\mu}^2\left(\bar{\Gamma}^2+\bar{\Omega}^2\right)}\right]\left(3\bar{\mu}^2\right)^{-1},
\end{eqnarray*}
The bistable transition occurs at
\begin{equation}\label{Input intensity for up-scan}
|\varepsilon_d|^2=\frac{f(n_{g1}^*)}{g_{mm}^2\left(\chi^2+\delta_l^2\right)}
\end{equation}
when the amplitude of the input field $\varepsilon_d$ is
up-scanned and for
\begin{equation}\label{Input intensity for down-scan}
|\varepsilon_d|^2=\frac{f(n_{g2}^*)}{g_{mm}^2\left(\chi^2+\delta_l^2\right)}
\end{equation}
when the amplitude of the input field is down-scanned (see
Fig.~\ref{Fig of bistable curves for phonon isolator}). For this
case too, we do not see a unidirectional phonon transport region
if we feed the inputs at the active and passive resonators sides
simultaneously. Thus we conclude that although the proposed system
can be used as phonon diode allowing nonreciprocal phonon
transport, it cannot function as an isolator for phonons.

\end{document}